\documentclass[fleqn,usenatbib]{mnras}
\usepackage{booktabs}
\usepackage{natbib}
\usepackage{hyperref}
\usepackage{footnote}
\usepackage{tablefootnote}
\usepackage{graphicx}
\usepackage{txfonts}
\usepackage{longtable}
\usepackage{url}
\usepackage{color}
\usepackage{subcaption}
\usepackage{float}
\usepackage{xcolor}

\newcommand{\teff}{T_{\rm eff}}
\newcommand{\logg}{\log g}
\newcommand{\vmic}{\xi_{\rm t}}
\newcommand{\vmac}{V_{\rm mac}}

\newcommand{\Elow}{E_{\rm low}}
\newcommand{\Eup}{E_{\rm up}}

\title[NLTE line formation of Y in cool stars]{Observational constraints on the origin of the elements. VII. NLTE analysis of Y II lines in spectra of cool stars and implications for Y as a Galactic chemical clock}

\author[N. Storm et al.]{Nicholas Storm,$^{1}$\thanks{0000-0002-5259-3974},Maria Bergemann,$^{1}$\thanks{0000-0002-9908-5571}
\\
$^{1}$Max-Planck Institute for Astronomy, K\"{o}nigstuhl 17, D-69117 Heidelberg, Germany\\
}

\begin{document}

\label{firstpage}
\pagerange{\pageref{firstpage}--\pageref{lastpage}}
\maketitle

\begin{abstract}

\noindent Yttrium (Y), a key s-process element, is commonly used in nucleosynthesis studies and as a Galactic chemical clock when combined with magnesium (Mg). We study the applicability of the previously assumed LTE line formation assumption in Y abundance studies of main-sequence and red giant stars, and probe the impact of NLTE effects on the [Y/Mg] ratio, a proposed stellar age indicator. We derive stellar parameters, ages, and NLTE abundances of Fe, Mg, and Y for 48 solar analogue stars from high-resolution  spectra acquired within the Gaia-ESO survey. For Y, we present a new NLTE atomic model. We determine a solar NLTE abundance of A(Y)$_{\rm NLTE}=2.12\pm0.04$ dex, $0.04$ dex higher than LTE. NLTE effects on Y abundance are modest for optical Y II lines, which are frequently used in Sun-like stars diagnostics. NLTE has a small impact on the [Y/Mg] ratio in such stars. For metal-poor red giants, NLTE effects on Y II lines are substantial, potentially exceeding $+0.5$ dex. For the Gaia/4MOST/WEAVE benchmark star, HD 122563, we find the NLTE abundance ratio of [Y/Fe]$_{\rm NLTE}=-0.55\pm0.04$ dex with consistent abundances obtained from different Y II lines. NLTE has a differential effect on Y abundance diagnostics in late-type stars. They notably affect Y II lines in red giants and very metal-poor stars, which are typical Galactic enrichment tracers of neutron-capture elements. For main-sequence stars, NLTE effects on optical diagnostic Y II lines remain minimal across metallicities. This affirms the [Y/Mg] ratio's reliability as a cosmochronometer for Sun-like stars.

\end{abstract}


\section{Introduction}

{Stellar age is a critical parameter for many fields in astronomy. The most common method for determining the age of a star is the method of fitting the stellar isochrones \citep[e.g.][]{Jorgensen2005, Gallart2005, Naylor2009, Schonrich2014, Kordopatis2023}, or evolutionary tracks \citep{Dotter2008, Serenelli2013, Gent2022}. Although this method works well for coeval samples of stars, such as open and globular clusters \citep{Naylor2006, Barker2018}, determinations of ages of field Galactic stars are more difficult because of parameter degeneracies \citep[e.g.][]{Pont2004, Serenelli2013}. Thus, recently several other methods have been explored in the literature. One of such methods is a so-called 'chemical clock' and it is based on empirical correlations between the ages of stars obtained using classical methods (or in combination with asteroseismology) and measured abundance ratios from stellar spectra \citep[see for example][]{Martig2016, Bergemann2016, Ness2016, SilvaAguirre2018, DelgadoMena2019, Skuladottir2019}.}

{The ratio [Y/Mg] has also been proposed as one such potential chemical clock \citep{daSilva2012}. Mg is produced mainly by the core-collapse supernovae explosions of massive stars with some contribution from intermediate-mass stars \citep{Timmes1995, Woosley1995, Thielemann2000, Kobayashi2006, Karakas2014}. Y is a slow neutron capture (s-process) element. Its production mainly happens in the intermediate-mass stars during their asymptotic giant branch (AGB) phase \citep{Busso1999, Cristallo2011, Karakas2016}. Owing to the differences in characteristic timescales of enrichment due to massive stars and AGB stars, the expectation is that [Y/Mg] should generally decrease with higher stellar age. Indeed, recently a well-defined correlation between [Y/Mg] and the stellar age was noted for solar twins in the local Galactic neighbourhood \citep{daSilva2012, Nissen2015, TucciMaia2016, Jofre2020, Berger2022}. The universality of the trend for stars in a larger Galactic volume was, however, questioned by \citet{Feltzing2017}, who found that the trend breaks down for lower metallicity stars, [Fe/H] $\lesssim -0.5$. The analysis of a large sample of stars by \citet{Casali2020} suggests that the [Y/Mg] and age relationship in not uniform with respect to different Galactocentric distances or star formation histories. This finding was attributed to the differences in star formation history of different Galactic populations coupled with metallicity dependence of s-process yields \citep[][]{Busso2001, Gallino2006, Karakas2016} leading to distinct [Y/Mg] trends.}

However, most of these studies assumed local thermal equilibrium (LTE) line formation for the Y abundance diagnostics. It has been well-established that departures from LTE, the NLTE framework, lead to significant differences in the abundance analysis. Specifically, for other s-process elements such as Sr and Ba, NLTE effects are known to be of the order of around 0-0.2 dex, and up to 0.5 dex for some Sr lines \citep{Bergemann2012b, Korotin2015, Gallagher2020}. It is therefore of interest to quantify the effects of NLTE in Y, in order to assess the consequences for standard diagnostics in the context of Galactic Chemical Evolution (GCE) \citep{Prantzos2018, Kobayashi2020} and chemical tagging \citep{Freeman2002, Hogg2016}.

{In this paper, we present a NLTE model atom of Y and apply the model in the analysis of spectra of Sun-like stars. In Sect. \ref{sec:obs_data}, we describe the observational data and the analysis of stellar parameters and ages. In Sect. \ref{sec:nlte_calc}, we provide the details on the statistical equilibrium calculations for Y and quantify the NLTE effects on the spectral lines of Y II commonly-used in the analyses of late-type stars. The abundance analysis of the solar-analogue sample is presented in Sect. \ref{sec:abund_determination}. The main results are described and compared to other literature studies in Sect. \ref{sec:results}. We close the paper with conclusions in Sect. \ref{sec:conclusion}.}
\section{Observations and stellar parameters}\label{sec:obs_data}

We use the optical high-resolution UVES spectra from the 4th public data release (DR) of the Gaia-ESO large spectroscopic survey\footnote{https://www.eso.org/qi/catalogQuery/index/393, http://archive.eso.org/scienceportal/home} \citep{Gilmore2022, Randich2022}. The spectra have the resolving power of $\lambda / \Delta \lambda \approx 47\,000$ and cover the optical wavelength range from 4770 to 6830 \AA~ with a small beam splitter gap at 5774-5833 \AA. 

Stellar parameters, including $\teff$, $\logg$, and [Fe/H], and ages of stars, were determined using the SAPP Bayesian pipeline (Stellar Abundances and atmospheric Parameters Pipeline) \citep{Gent2022}. The code adopts NLTE spectroscopic models as described in \citet{Kovalev2019} and the GARSTEC evolutionary tracks \citep{Weiss2008}. In addition to spectra, we also include Gaia DR3 photo-geometric distances \citep{BailerJones2021}, as well Gaia DR3 \citep{Gaia2022}, 2MASS \citep{Skrutskie2006}, and APASS \citep{Henden2015} photometry. The Bayesian analysis of stellar parameters closely follows the method outlined in \citet{Gent2022} {(see their Sec. 3.9 for the full procedure)}, with the only exception that the asteroseismic module is not used, since for none of our solar analogues global oscillation quantities are available. {The main advantage of the SAPP code is in its ability to combine all observational information available for stars, including spectra, photometry, and astrometry, in order to provide homogeneous estimates of fundamental stellar parameter in the same self-consistent comprehensive statistical framework. As such, the code does not assume the spectroscopic estimates to perform the analysis of ages, but rather provides $\teff$, $\logg$, [Fe/H], masses, and ages of stars from the combination of spectroscopic and photometric probability distribution functions (PDF), the former assisted by grids of stellar spectra and the latter involving models of stellar evolution and parallax \citep[for a similar, but not identical  approach, see also][]{Pont2004, Jorgensen2005, Serenelli2013, Schonrich2014}. The result is the posterior PDF in the multi-dimensional parameter space, with the uncertainties reflecting the true shape of the individual distribution functions corresponding to individual observables. The final estimates of stellar parameters are computed from the moments of the posterior PDF. For the complete mathematical and numerical framework we refer the reader to \citet{Gent2022}.}

\begin{figure}
\includegraphics[width=1\columnwidth]{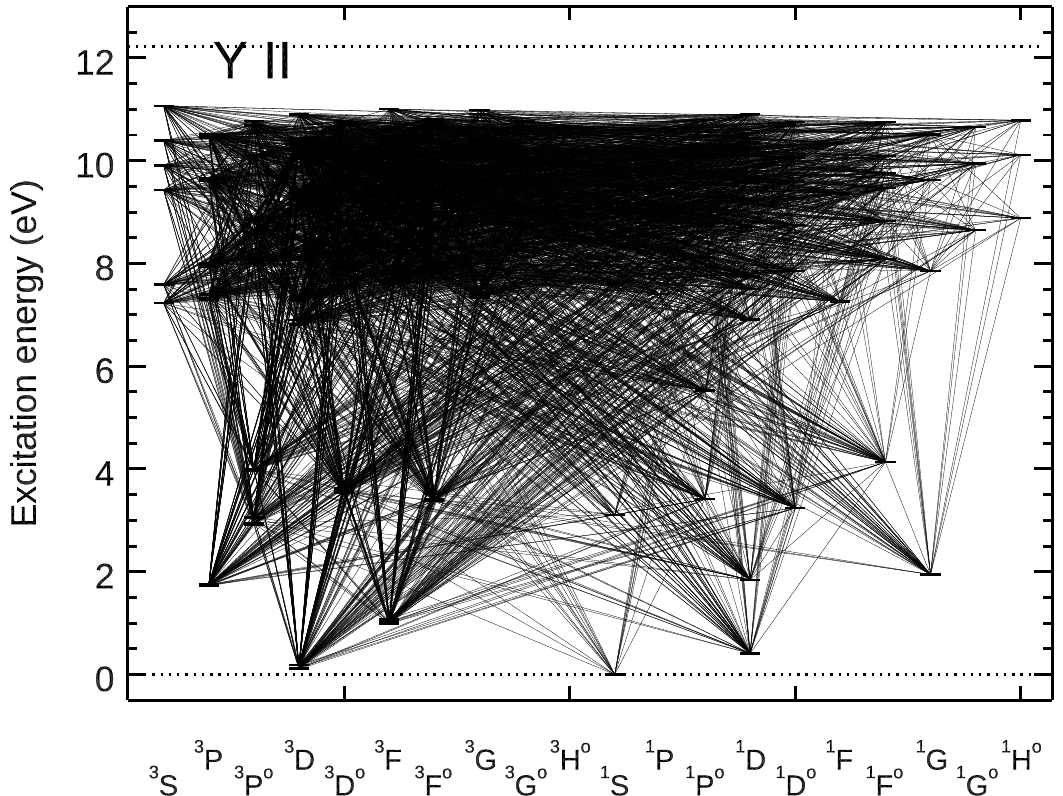}
\includegraphics[width=1\columnwidth]{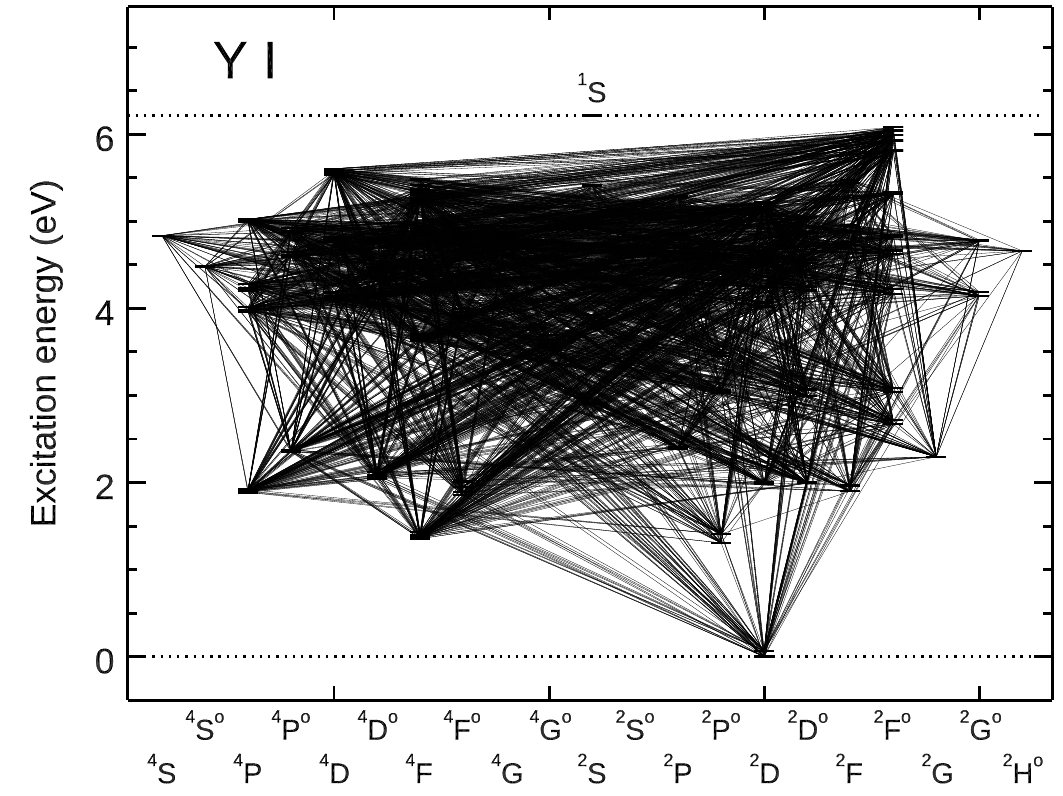}
\caption{Grotrian diagrams of the NLTE Y model. The model comprises 424 energy states representing Y I, Y II, and Y III ionisation stages. The ionisation potentials of Y I and Y II are $6.22 $ eV and $12.22$ eV, respectively.}
 \label{fig:grotrian}
\end{figure}
\begin{figure*}
\hbox{
\includegraphics[width=0.48\textwidth]{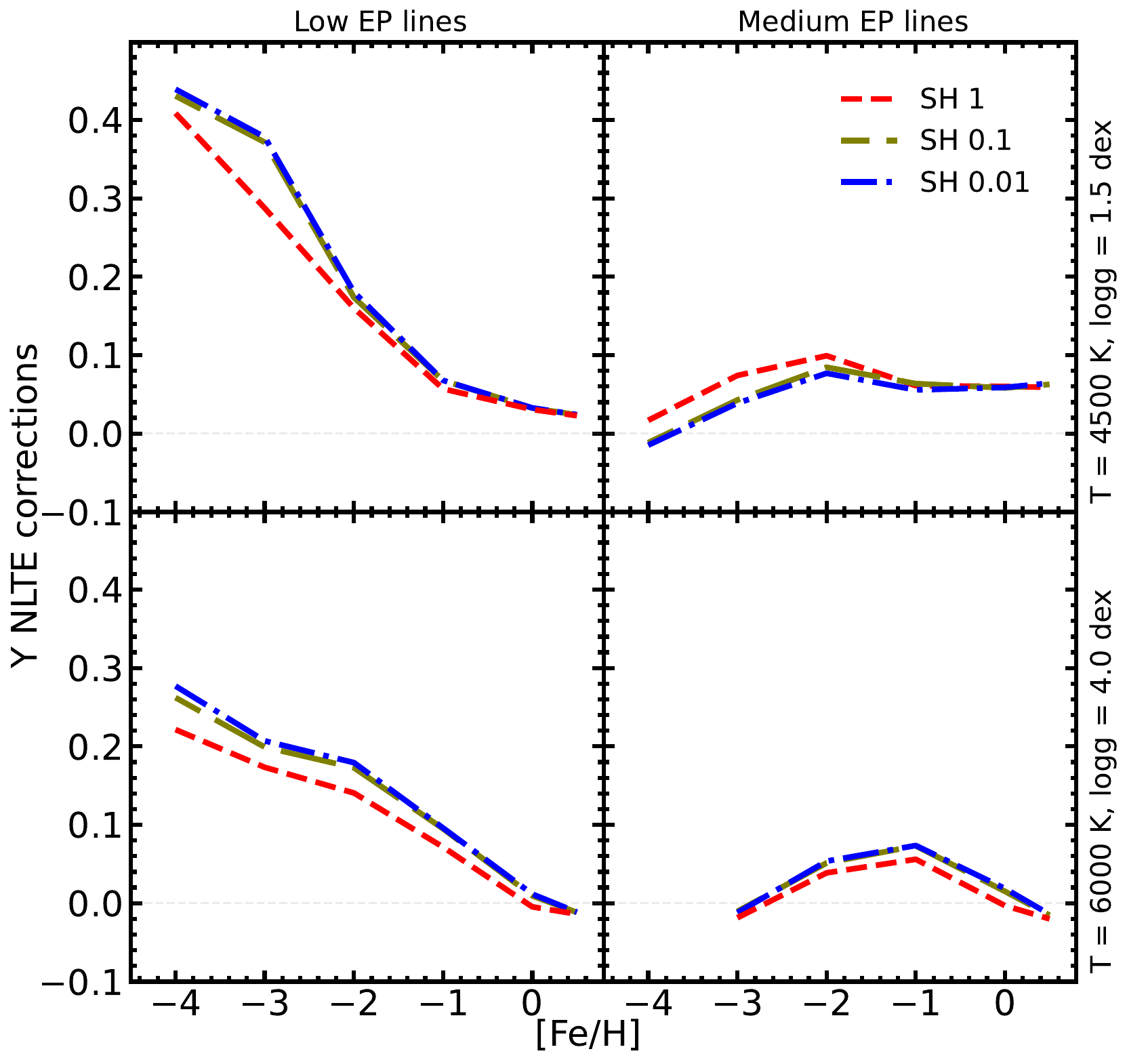}
\includegraphics[width=0.48\textwidth]{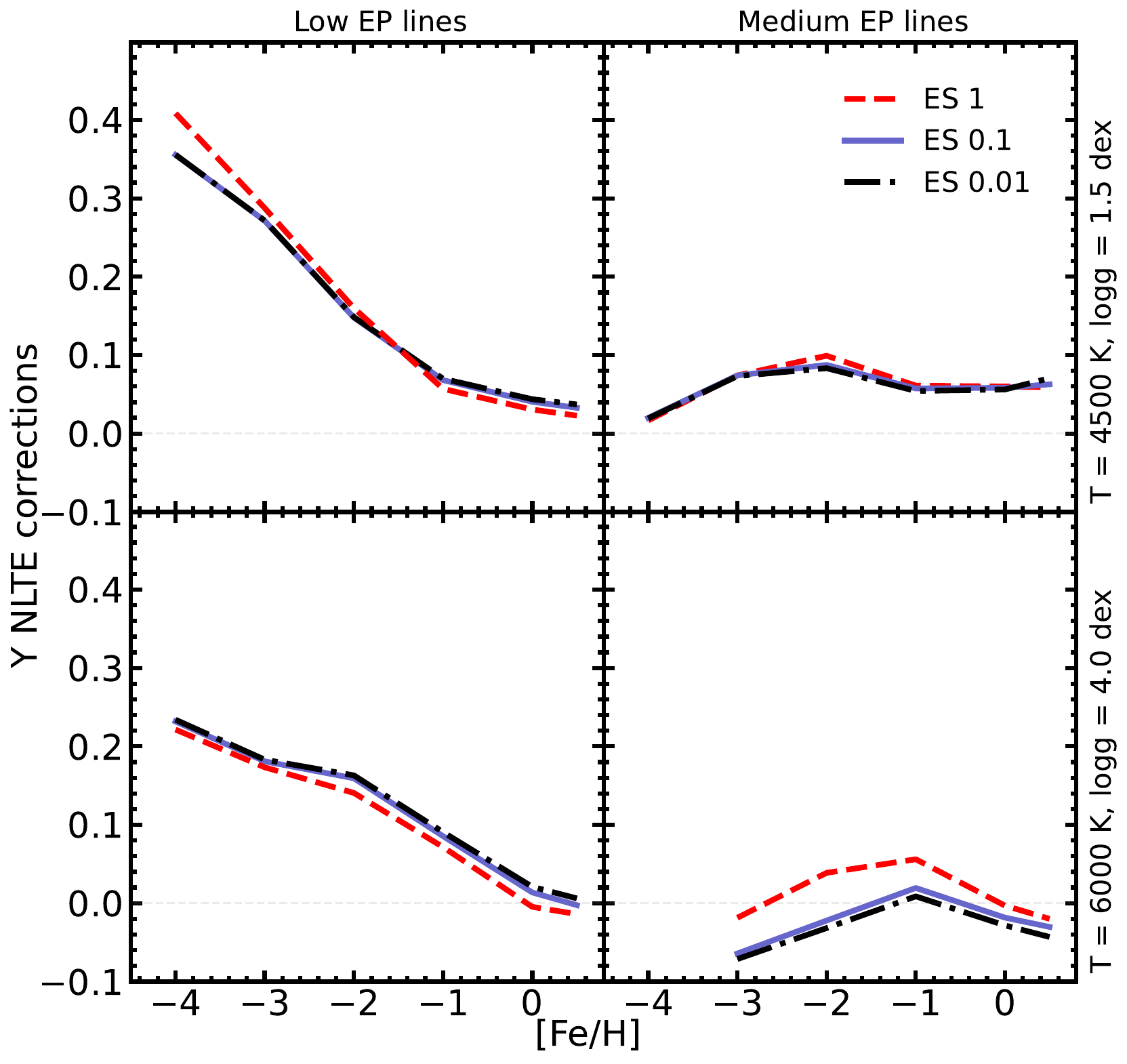}}
\caption{Impact of using different scaling factors with H or with electrons. Representative RGB and main-sequence are used for low and medium excitation potential lines. {Our reference model atom assumes standard unscaled rate coefficients for transitions in Y$+$H and Y$+$e$-$ inelastic collision reactions that corresponds to scaling factors set to SH $=1$ and ES $=1$.}}
 \label{fig:scaled_sh_impact}
\end{figure*}

Since the scope of this work is to test the Y/Mg relation with age for solar analogues, out of the entire high-resolution Gaia-ESO sample, we have selected $48$ stars that closely match the atmospheric parameters of the Sun, that is $\teff = 5777\pm 100$ K and $\logg = 4.44\pm 0.1$ dex. This sample comprises stars within a range of ages from $4$ to $11$ Gyr and heliocentric distances from $0.4$ to $1.6$ kpc. We did not impose any restrictions on [Fe/H], however, the choice on $\teff$ and $\log g$ naturally limits the sample to stars with [Fe/H] from $\sim -0.6$ to $+0.1$ dex. The spectral signal-to-noise ratio (SNR) ranges from 24 to 130 with a median of 48 per pixel. The errors of $\teff$, $\logg$, and [Fe/H] are on average $132$ K, $0.1$ dex, and $0.01$ dex, respectively. Ages are accurate to $\sim 30 \%$ \citep[see also][]{Serenelli2013}, which is fully sufficient for the purposes of this paper, as we are not aiming to define an age-abundance relationship but only to investigate the impact of NLTE effects in the Y/Mg ratio regardless of the nature and slope of the correlation with age. 

We note that, not only to enable a more comprehensive analysis of LTE and NLTE trends, but also since the model atom of Fe has been improved as discussed in \citet{Semenova2020}, we have chosen to re-derive metallicities in LTE and in NLTE using the same approach as that adopted for Y and Mg, as described in Sect. \ref{sec:abund_determination} below. Comparison of the new NLTE metallicities with the NLTE SAPP values suggests a very good agreement between the metallicity scales with the mean difference of only $0.05$ dex.

\section{NLTE analysis}\label{sec:nlte_calc}

The Y model atom was constructed using data from the Kurucz database\footnote{http://kurucz.harvard.edu/atoms/3900/}. The models contains 423 Y I and Y II levels, and it is closed by the ground state of Y III. This element only has one stable isotope, $^{89}$Y and the hyperfine splitting is very small \citep[][]{Hannaford1982,Nissen2017}, with the splitting less than 1 m\AA, and thus can be neglected in the line formation calculations. The model includes $11\,819$ radiative bound-bound transitions with the oscillator strengths from \citet{Biemont2011}. The lines span the entire wavelength range from $1\,000$ \AA$\textrm{ }< \lambda < 100\,000$ \AA. The radiative bound-free transitions are calculated using the hydrogenic recipe; this is not a significant source of uncertainty, because of the very large ionisation potential of Y II of $12$ eV, which makes radiative ionisation (the main source of NLTE effects in neutral ions, \citealt{Bergemann2014}) extremely inefficient similar to other singly-ionised species, such as Sr \citep{Bergemann2012c} and Ba \citep{Gallagher2020}. For the lack of quantum-mechanical data, collisions with e$-$ and H atoms are described using the standard formulae from \citet{Regemorter1962}, \citet{Seaton1962}, and \citet{Drawin1968}. To check the sensitivity of the NLTE corrections to collisional data, we performed additional calculations with scaling factors applied systematically to all rate coefficients, corresponding to the Y$+$H and Y$+$e- reactions. The results are shown in Fig. \ref{fig:scaled_sh_impact}. Clearly, decreasing the Drawin's rates for bound-bound and bound-free transitions by a factor of 10 (SH $=0.1$) or 100 (SH $=0.01$) does not lead to any significant changes in the NLTE corrections, although, as expected the corrections {slightly change, by $\sim +0.03$ to $+0.05$ dex at lower metallicity}. Likewise, the variation of the scaling factor to electron collision rates, ES $=0.1, 0.01$ leads to nearly identical results. The maximum change does not exceed $0.04$ dex in metal-poor main-sequence model atmospheres. We note, however, that these classical collisional formulae typically over-estimate the cross-sections of reactions associated with particle-particle collisions by several orders of magnitude \citep{Barklem2016, Bergemann2019, Bergemann2021} and consequently our results obtained with the reference model atom, which assumes un-scaled rate coefficients (SH $=1$ and ES $=1$), can be viewed as conservative estimates of the NLTE effects in the diagnostic Y II lines.

The statistical equilibrium (SE) calculations are performed the MULTI1D code, which solves radiation transfer equation in one-dimensional (1D) plane-parallel geometry and the SE equations in the iterative process. The code was originally published by \citet{Carlsson1986} and further updated by different research groups over the years to match the individual scientific needs. The updates performed in our group are described in \citet{Bergemann2019} and \citet{Gallagher2020}. We make use of the MARCS 1D LTE model atmospheres \citep{Gustafsson2008} that are available for {the  parameter space of late A and FGKM-type stars} ($2500 \leq \teff \leq 8000$ K), surface gravity ($-1 \leq \logg \leq 5$ dex), metallicity ($-5 \leq \textrm{[Fe/H]} \leq +1$ dex) and microturbulence ($\vmic = 0, 1, 2, 5$ km/s, where different temperature and surface gravity models have different available parameters chosen from these 4 options). The solar-metallicity opacity file, which contains the absorption properties of various atomic species, ions, and molecules in the atmosphere (referred to as ABSMET in MULTU1D), was used for all model calculations. This is because using opacity files with different metallicities had no impact on the line profiles of any diagnostic Y II lines. The departure coefficients for the MARCS grids were calculated using a Python wrapper to MULTI1D \footnote{\url{https://github.com/stormnick/wrapper_multi}} and these are available on our public database that also includes other NLTE grids for the elements (Ba, Ca, Co, Fe, H, Mg, Mn, Na, Ni, O, Si, Sr, Ti) at \url{https://keeper.mpdl.mpg.de/d/6c2033ef5c5d4c9ca8d1/}.

The selection of diagnostic Fe, Mg, and Y lines is carried out on the basis of the Gaia-ESO linelist \citep{Heiter2021}, which represents a careful compilation of atomic and molecular data used in the Gaia-ESO survey analysis. The linelist was recently updated with new atomic data for several elements \citep{Magg2022}, including Mg, for which the oscillator strengths from \citet{PehlivanRhodin2017} are used. The list of Fe lines comprises 52 Fe I and 8 Fe II lines with YY, YU, or UY flags in the Gaia-ESO linelist. These lines were selected through an initial fit applied to a larger sample of Fe lines. However, through a careful visual inspection we then removed the strong lines, lines resulting in poor fits (e.g. because of blends), or lines yielding vastly inconsistent abundances (possibly because of inaccurate $gf$-values). For Y II, the $\log gf$ values are taken from \citet{Biemont2011}, {who estimated the transition probabilities by combining the new level lifetimes measured using the technique of laser induced fluorescence with the theoretical branching fractions computed by means of pseudo-relativistic Hartree-Fock method with core-polarization effects. In \citet{Biemont2011}, no errors are provided for individual transitions, however, the authors note that the uncertainties are expected to range from 5 to 15 $\%$, with errors being larger for weaker transitions. Most of our diagnostic lines of Y II are rather strong, as judged by the laboratory intensity value in col. 3 of Table 5 in \citet{Biemont2011}. Therefore, here we conservatively adopt the uncertainty of 0.05 dex ($\approx 11 \%$) for Y II transitions, except the transitions at 5289 and 5728 \AA, for which we adopt the oscillator strength error of 0.07 dex ($\approx 15 \%$). Damping caused by elastic collisions with H atoms follows the quantum-mechanical data from \citet{Barklem2000}, for the 5711 \AA~Mg I line we employ the $\gamma$ and $\sigma$ values from P. Barklem (priv. comm). If these are not available, as is the case for the Y II lines, the cross-sections are calculated from the theory of \citet{Anstee1991, Anstee1995}.} 

Through a careful inspection of the line profiles, we found that out of the four diagnostic Y II lines in the Gaia-ESO UVES data only the line at $4883$ \AA~ is sufficiently strong and clearly discernible from the other features in all spectra in our stellar sample. Therefore the final abundance analysis of the solar analogues we primarily rely on this Y II feature. However, the solar abundance analysis is carried out using a larger set of Y II lines. Atomic data {for the diagnostic Y II and Mg I lines are provided in Table \ref{tab:lines}.} 

\begin{table}
\begin{minipage}{\linewidth}
\renewcommand{\footnoterule}{} 
\setlength{\tabcolsep}{1.5pt}
\caption{Atomic parameters for the diagnostic spectral lines of Mg I and Y II. {See text.}}
\label{tab:lines}     
\begin{center}
\begin{tabular}{l c cc ccc}
\noalign{\smallskip}\hline\noalign{\smallskip}  $\lambda$ & $\Elow$ & $\Eup$ & $\log gf$ & vdW\footnote{Van der Waals broadening parameter, see text.} & Ref.\footnote{References: ~~~
(1) \citet{PehlivanRhodin2017}
(2) \citet{Biemont2011}
}

& Ref.\footnote{References: ~~~(3) \citet{Barklem2000} (4) P. Barklem (priv. comm.)  (5) \citet{Kurucz2011}}  \\
   ~~~~[\AA] & [eV] & [eV] & & & f-val. & vdW \\
\noalign{\smallskip}\hline\noalign{\smallskip}

Mg I  & & & & & & \\
5528.405  &  4.346  & 6.588  & $-0.55 \pm 0.02$ & 1461.312  & 1 & 3 \\ 
5711.088  &  4.346  & 6.516  & $-1.74 \pm 0.05$ & 1860.100  & 1 & 4 \\ 

Y II & & & & & & \\
4883.682 & 1.084 & 3.622 & {~~$0.19 \pm 0.05$} & -7.760 & 2 & 5 \\ 
4900.118 & 1.033 & 3.562 & {~~$0.03 \pm 0.05$} & -7.760 & 2 & 5 \\
5087.416 & 1.084 & 3.520 & {$-0.16 \pm 0.05$} & -7.810 & 2 & 5 \\ 
5200.406 & 0.992 & 3.376 & {$-0.47 \pm 0.05$} & -7.800 & 2 & 5 \\
5289.815 & 1.033 & 3.376 & {$-1.68 \pm 0.07$} & -7.800 & 2 & 5 \\ 
5728.887 & 1.839 & 4.003 & {$-1.15 \pm 0.07$} & -7.750 & 2 & 5 \\ 

\noalign{\smallskip}\hline\noalign{\smallskip}
\end{tabular}
\end{center}
\end{minipage}
\end{table}

\section{Abundance analysis}
\label{sec:abund_determination}

For the detailed abundance analysis, we use the Turbospectrum-NLTE (also known as TSv20) code \citep{Gerber2023}, which is an updated version of the original Turbospectrum code \citep{Alvarez1998, Plez2012}. The code is open-source\footnote{\url{https://github.com/bertrandplez/Turbospectrum\_NLTE}} and allows computations of synthetic stellar spectra with NLTE effects for multiple chemical elements at once. To find the best-fit abundances, we make use of the TSFitPy wrapper \citep{Gerber2023} that relies on the Nelder-Mead optimisation. For the purpose of this work, TSFitPy code was refactored, optimised and several new features were added. One of the new features is the ability to fit macroturbulence ($\vmac$) and rotation for individual spectral lines. This is done using the L-BFGS-B (Limited-memory Broyden-Fletcher-Goldfarb-Shanno with bound consideration) algorithm \citep{Byrd1995, Zhu1997} as a secondary step after fitting an abundance to break the degeneracy of fitting these two parameters. Additionally, the code was parallelized using the Dask Python package \citep{Rocklin2015, Dask2016}.

The abundances were determined via a strictly differential analysis, {that is, relating the abundance from each individual line to the measurement from the solar spectrum,}  with respect to the solar UVES spectra from the Gaia-ESO benchmark library \citep{BlancoCuaresma2014}. One of the main advantages of the differential analysis is that one can minimise the errors caused by, e.g. atomic and molecular data, and  for stars with very similar atmospheric parameters, this approach allows to partly counterbalance the physical limitations of the model atmospheres and line formation, such as the lack of 3D convection \citep[e.g.][]{Nissen2018}. For the solar UVES spectrum, we obtain in NLTE, A(Fe) $= 7.53$ dex and A(Mg) $= 7.55$ dex, respectively. These estimates are fully consistent with the results by \citet{Bergemann2012a}, \citet{Bergemann2017}, and \citet{Magg2022}, respectively. The detailed solar results for Y II will be discussed in detail in the next section.

\begin{figure*}
\hbox{
\includegraphics[width=0.48\textwidth]{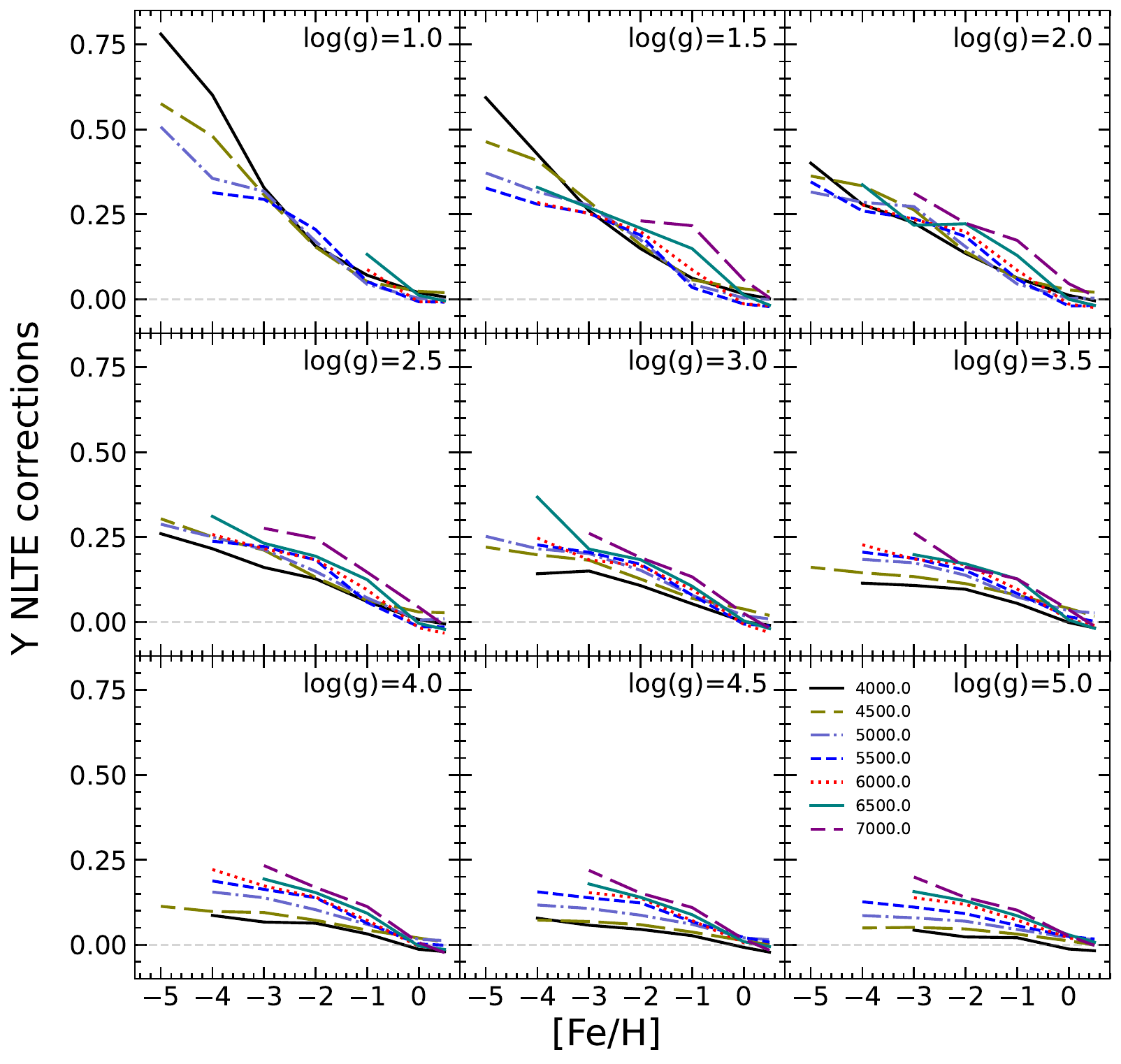}
\includegraphics[width=0.48\textwidth]{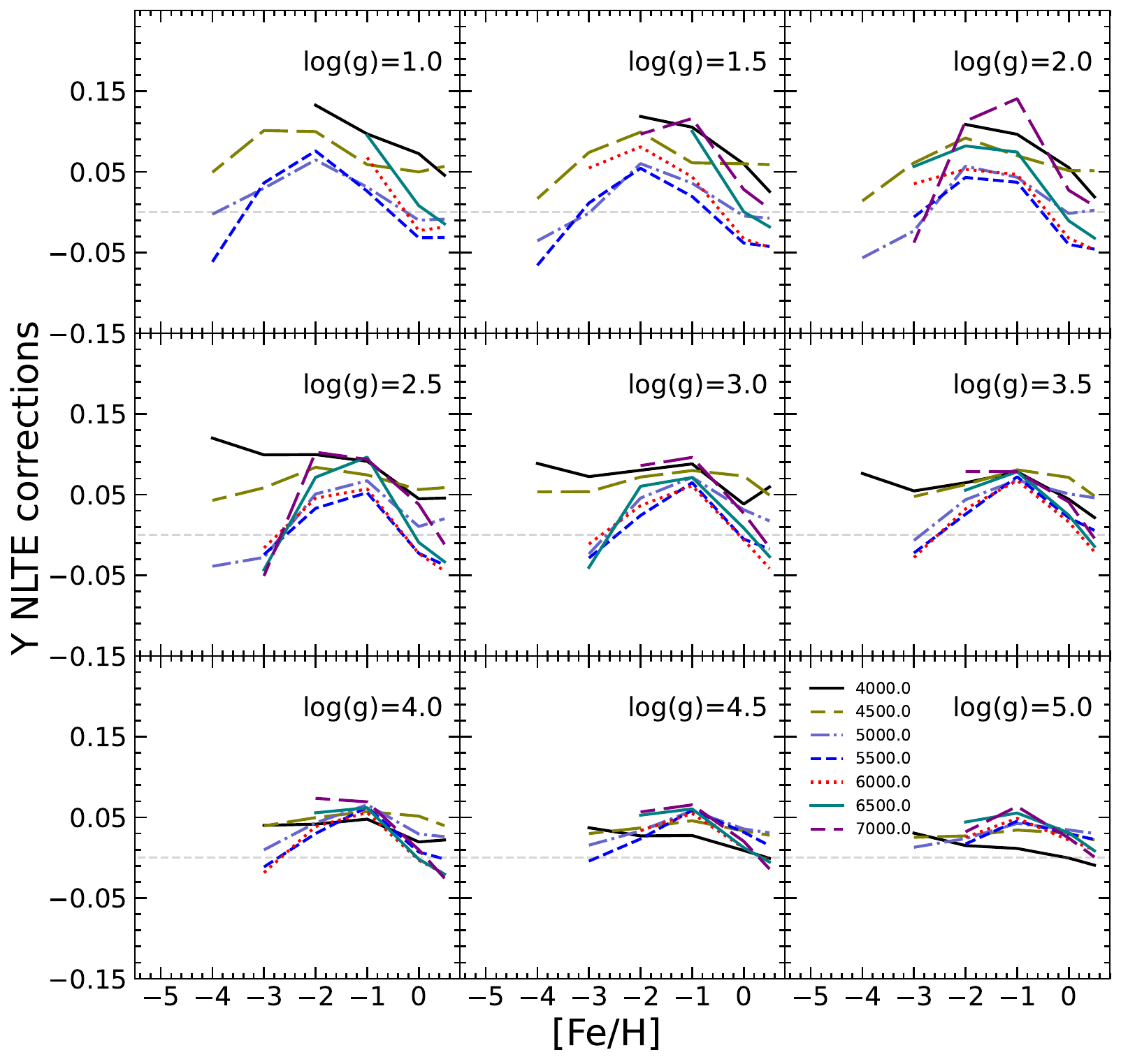}}
\caption{Average NLTE corrections for the low-EP (multiplets 7,8,6) and medium-EP (multiplets 21, 23) Y II lines plotted against metallicity [Fe/H] of the model atmospheres. We adopt $\vmic = 2$ km/s and $\vmic = 1$ km/s for the models with log(g)$ \leq 3.5$ dex and log(g)$ > 3.5$ dex respectively.}
\label{fig:nltecor}
\end{figure*}

{To estimate the systematic abundance error, we follow the approach adopted by \citet{Bergemann2017}, in which the individual stellar parameters are varied according to their respective uncertainties and the change in the measured abundance ratios is derived by adding individual errors in quadrature:}
{
\begin{equation}
    \sigma_{\rm{syst}} = \sqrt{\sigma^2_{\teff} + \sigma^2_{\logg} + \sigma^2_{\rm{[Fe/H]}}},
\end{equation}
}
\noindent {using average errors on stellar parameters described in Sec. \ref{sec:obs_data}. This procedure yields systematic uncertainties $\sigma_{\rm{A(Y)}} = 0.04$ dex and $\sigma_{\rm{A(Mg)}} = 0.09$ dex. We also note that for the Sun, the primary source of the uncertainty in the abundance estimate is the quality of the oscillator strengths and physical limitations of models. This will be addressed in more detail in Sect. \ref{subsec:the_sun}.} 
\section{Results}\label{sec:results}
{We begin the analysis of results with the discussion of NLTE effects in lines of Y II. To allow for a more comprehensive understanding of NLTE effects, we also include a wider range of Y II lines from the near-UV to the optical (Table \ref{tab:nlte} in Appendix), in order to make a more direct reference to studies of Y abundances in FGK-type stars in the literature.}

{We also limit the analysis to lines with the EW greater than $1$ m\AA, as weaker features are not detectable in our observed spectra. The NLTE effects are quantified via the standard concept of the NLTE abundance correction \citep{Bergemann2014}, which represents the difference in abundance that is required to match the equivalent width (EW) of a NLTE model line to that of the LTE line computed using the identical values of stellar parameters and Y abundance.}
\subsection{NLTE effects on Y II lines}
Figure \ref{fig:nltecor} shows the average NLTE corrections for two groups of Y II lines. The lines are grouped by the excitation potential (EP) of their lower energy state, such that the low-EP features are those that originate from states with $\sim 0.1$ eV and the corresponding spectral lines are visible in the near-UV and blue range of FGK-type stars, such as the lines at 3710.29, 3774.33 and 4398.01 \AA. The medium-EP lines arise from the energy states around $\sim 1$ eV, leading to a set of rather weak spectral lines in the optical; these are sometimes employed in the solar abundance analysis \citep{Grevesse2015}. Some representative Y II lines of the latter group are the lines at 4883.68, 4900.12, and 5087.42 \AA.

As seen in Fig. \ref{fig:nltecor}, overall the NLTE corrections to Y II lines are typically positive and they increase with decreasing metallicity or increasing the $\teff$ of the model atmosphere. This is the consequence of progressively stronger UV radiation fields in more metal-poor, or hotter, environments that enhances the over-excitation in the strong Y II lines in the UV and in the optical wavelength ranges. At the solar metallicity, we find NLTE effects at the level of only $\sim \pm 0.05$ dex for all models atmospheres and all Y II lines. For the main-sequence and turn-off stars, the NLTE corrections for the low-EP lines typically do not exceed  $+0.2$ dex even at [Fe/H] $\approx -4$. However, for red giant models, NLTE corrections may reach up to $+0.75$ dex at [Fe/H] $\approx -5$. For the medium-excitation Y II lines, such as those of multiplets 21 and 23, the NLTE corrections are typically much more modest and do not exceed $0.05$ dex for the main-sequence stars and $0.15$ dex for red giant models even at very low [Fe/H]. Interestingly, the change of slope of the NLTE correction is usually seen at [Fe/H] $\sim -1$, which may have an impact on the GCE trend of [Y/Fe] \citep{Kobayashi2020} or, for example, on the ratio of [Ba/Y] which is used to quantify the s-process efficiency \citep{Travaglio2004, Prantzos2018}.

\begin{figure*}
\hbox{
\centering
\includegraphics[width=0.45\textwidth]{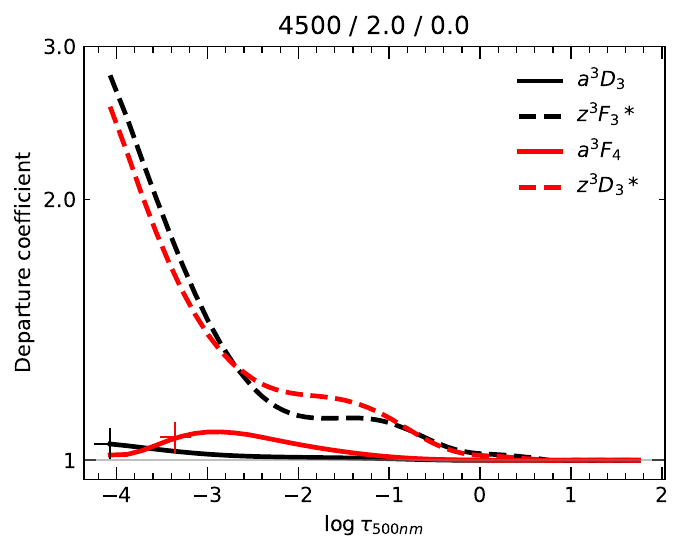}
\includegraphics[width=0.45\textwidth]{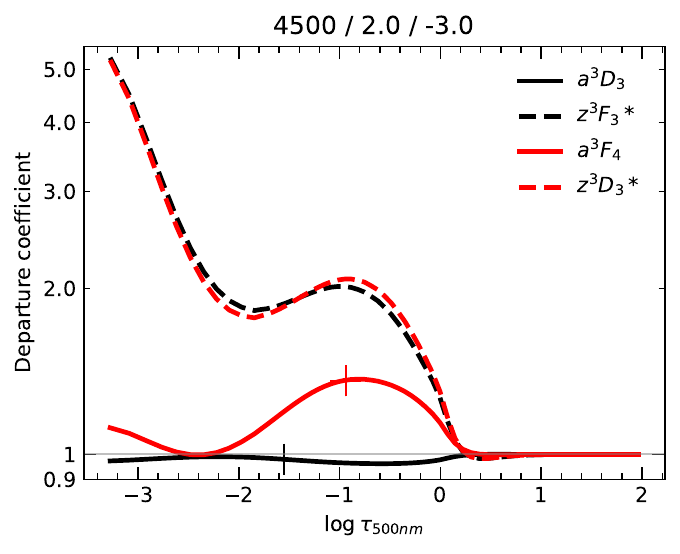}
}
\hbox{
\centering
\includegraphics[width=0.45\textwidth]{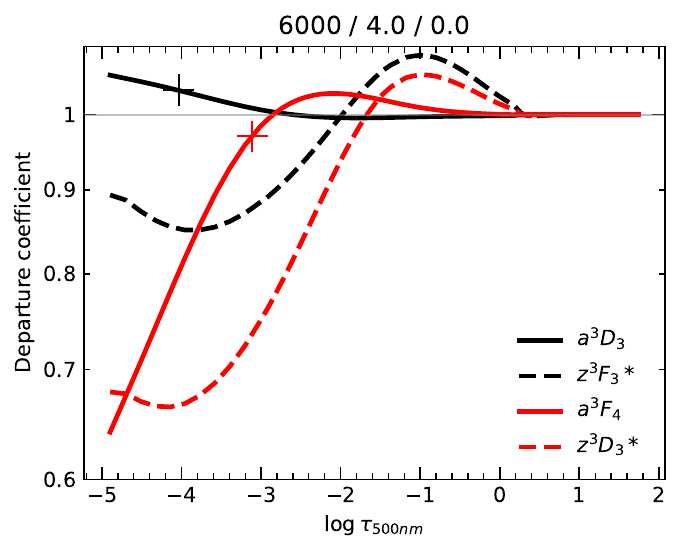}
\includegraphics[width=0.45\textwidth]{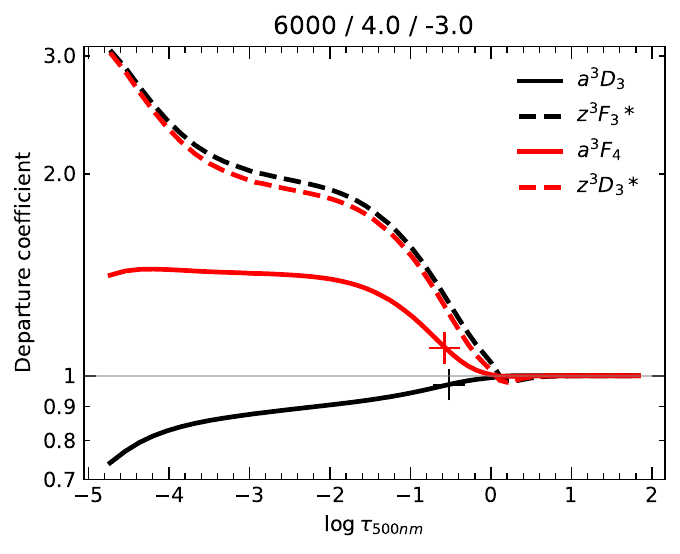}
}
\caption{{Departure coefficients as a function of model atmosphere's optical depth for a red giant model on top ($\teff = 4500$ K, $\logg = 2.0$ dex for [Fe/H] = -3 and 0, respectively left and right plots) and a main-sequence on the bottom ($\teff = 4500$ K, $\logg = 2.0$ dex for [Fe/H] = -3 and 0, respectively left and right plots). Low-EP line (3832 \AA) levels (a $^3$D$_3$ and z $^3$F$_3$*) are depicted in black and medium-EP line (4883 \AA) levels (a $^3$F$_4$ and z $^3$D$_3$*) in red, with solid and dashed lines representing lower and upper levels, respectively. Vertical lines indicate line core at $\log \tau = 0$ for both low-EP (black) and medium-EP (red) lines.}}
\label{fig:depcoef_rg}
\end{figure*}

{To further examine the differences between the line formation of low- and medium-EP lines, we show the departure coefficients $b_i$ as a function of optical depth in Fig. \ref{fig:depcoef_rg}. Here $ b_i = b\frac{n_i \rm NLTE}{ n_i \rm LTE}$, i.e. the quantity represents the ratio of NLTE to LTE populations of the energy state $i$. We include four limiting cases: a red giant model (top) and a main-sequence model (bottom) with metallicities [Fe/H] $= 0$ and $-3$, in order to visualise the variations in line formation and their dependencies on both excitation potential and stellar metallicity. The black lines correspond to low-EP (3832 \AA), while red ones correspond to medium-EP (4883 \AA) lines. The formation heights of the cores of the diagnostic Y II lines ($\log \tau = 0$ for line core) are shown with vertical lines.}

{Let us first focus our attention on the top left panel in Fig. \ref{fig:depcoef_rg}, which shows the solar metallicity, [Fe/H]$=0$, model atmosphere of a red giant star. Both low- and medium-EP lines form in the outermost parts of the atmospheres, at $\log_{\rm \tau 500} \lesssim -3$. In these layers, the lower energy states of both lines, a$^3$D and a$^3$F, tend to be close to unity (b$_i \approx 1$). However their upper energy states $j$, z$^3$F$*$ and z$^3$D$*$, are significantly over-populated with respect to the LTE values. Following \citet{Bergemann2014}, this implies that the line source function exceeds the Planck function $S^{\rm l}/B^{\rm l}_{\nu} \approx b_j/ b_i > 1$. This results in line weakening under NLTE conditions compared to LTE and slightly positive NLTE corrections.}

At low metallicity, [Fe/H]$=-3$ dex, another effect becomes relevant. Owing to less line blanketing, the more intense radiation field in the Wien's regime leads to stronger radiative pumping in the transitions and correspondingly larger over-populations of higher energy states in Y II. This is manifested in an even larger source function vs Planck function imbalance, compared to the solar metallicity. But we also find a noticeable over-population of the medium-EP states, such as the a$^3$F level, which represents the lower level of the medium-EP line. Consequently, the line opacity is slightly enhanced compared to LTE ($\kappa^{\rm l} \sim b_i > 1$), and the NLTE corrections decrease, also they remain positive. This effect of opacity is further amplified as metallicity decreases, and hence the NLTE corrections for the medium-EP lines change sign, eventually becoming negative at very low [Fe/H]. In contrast, the a$^3$D state of Y II remains nearly thermalised also at low metallicity, $\kappa^{\rm l} \approx 1$, and the progressively increasing $S^{\rm l}/B^{\rm l}_{\nu}$ imbalance leads to steadily increasingly positive NLTE corrections for the low-EP Y II lines over the entire metallicity range.

For the main-sequence model atmospheres, the $b_i$ factors show a qualitatively different behaviour. At solar [Fe/H], the lower energy states of low- and medium-EP  transitions remain close to unity throughout the entire range of optical depths, $-3 \lesssim \log_{\rm \tau 500} \lesssim 1$. The ground state of the low-EP line is nearly thermalised at the depth of the core formation, which implies LTE line opacity. Although the line source function is slightly sub-thermal (as $b_j/b_i <1$), the Y II line is strong and the NLTE correction, although negative, remains close to zero. For the medium-EP line, the somewhat larger effect on $S^{\rm l}$ is counter-balanced by the weak depopulation of the lower energy state a$^3$F, implying smaller opacity in NLTE. Also in this case, the net effect is that the NLTE correction is about zero. At low metallicity, [Fe/H] $=-3$, the Y II lines become very weak and the NLTE corrections follow the same trend as that observed for low-[Fe/H] RGB models. The low-EP line is weakened under NLTE conditions, owing to  decreased line opacity and increasing NLTE corrections. In contrast, the NLTE effects on the medium-EP line decrease, due to the progressively increasing over-population of its energy state.

\begin{figure}
\includegraphics[width=1\columnwidth]{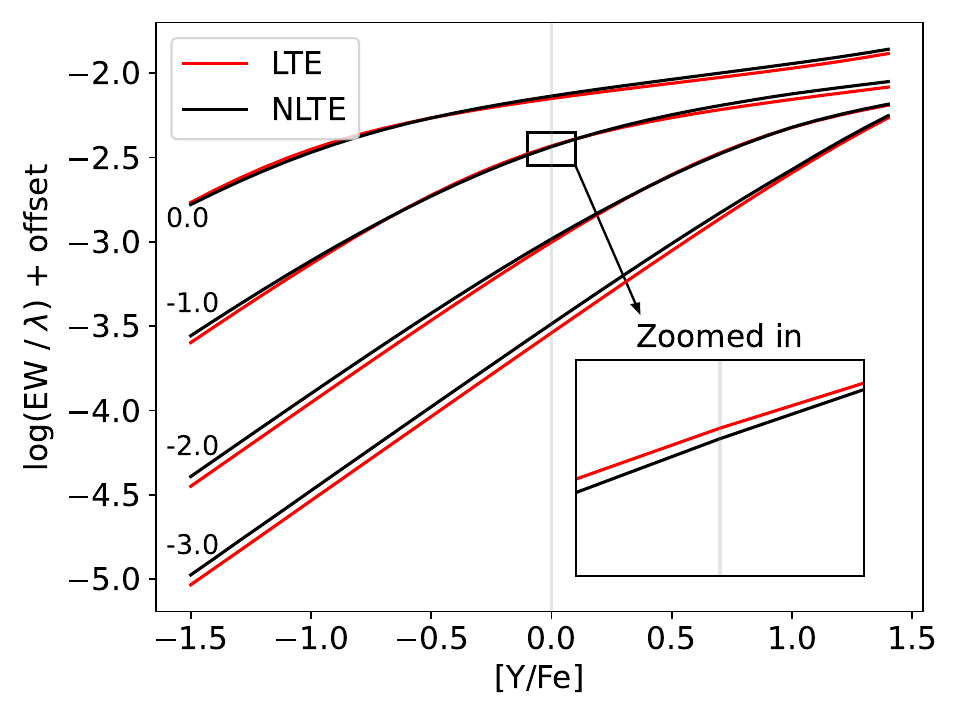}
\caption{LTE (in red) and NLTE (in black) curve of growth for a model atmosphere log(g) = 3 dex and $\teff$ = 6500 K for metallicities from -3 to 0 (labelled next to each curve) for a medium-EP line. At [Fe/H] = -1 CoG transitions from linear to saturated one.}
 \label{fig:cog}
\end{figure}

{To further investigate the change of slope for the medium-EP lines at [Fe/H]~$\approx-1$, in Fig. \ref{fig:cog} we show the curve-of-growth (CoG) for the Y II line 4883 \AA~for the model atmosphere of a star with $\teff = 6500$ K and $\logg=3$ and different metallicities.} The NLTE CoG is plotted in black, while the LTE CoG is in red. Since Fig. \ref{fig:nltecor} was computed for [Y/Fe] = 0, we focus our attention on that part of the graph as well. For [Fe/H]~$ \lesssim -3$, the Y II line is clearly on the linear part of the CoG, unless an extreme [Y/Fe] enhancement is adopted. However, at around [Fe/H] $\approx -1$ (the exact {value} is a bit different for each line), the line undergoes a transition from the linear to saturated part of the CoG, and for higher metallicities this line is primarily saturated, even for sub-solar [Y/Fe] ratios. {At this transition point, the LTE line becomes stronger leading to positive NLTE corrections. On the other hand, for other metallicity values, the LTE line is instead weaker giving rise to negative NLTE corrections.} This curious behaviour results in a peak for NLTE corrections around [Fe/H]~$\approx -1$ as seen in the right part of the Fig. \ref{fig:nltecor}.

To summarise, we find that the NLTE has a different effect on the near-UV and optical lines of Y II. The lines emerging from the lowest excitation energy states tend to be weaker when calculated in NLTE, and the effect increases with decreasing [Fe/H] or increasing $\teff$ of the model atmosphere. In contract, the lines emerging from the energy states above $\sim 1$ eV are not so sensitive to NLTE, however, modest effects at the level of $+0.1$ dex can be expected depending on the metallicity, $\teff$, and $\logg$ of the star.
\subsection{The Sun}
\label{subsec:the_sun}
\begin{figure}
\includegraphics[width=1.\columnwidth]{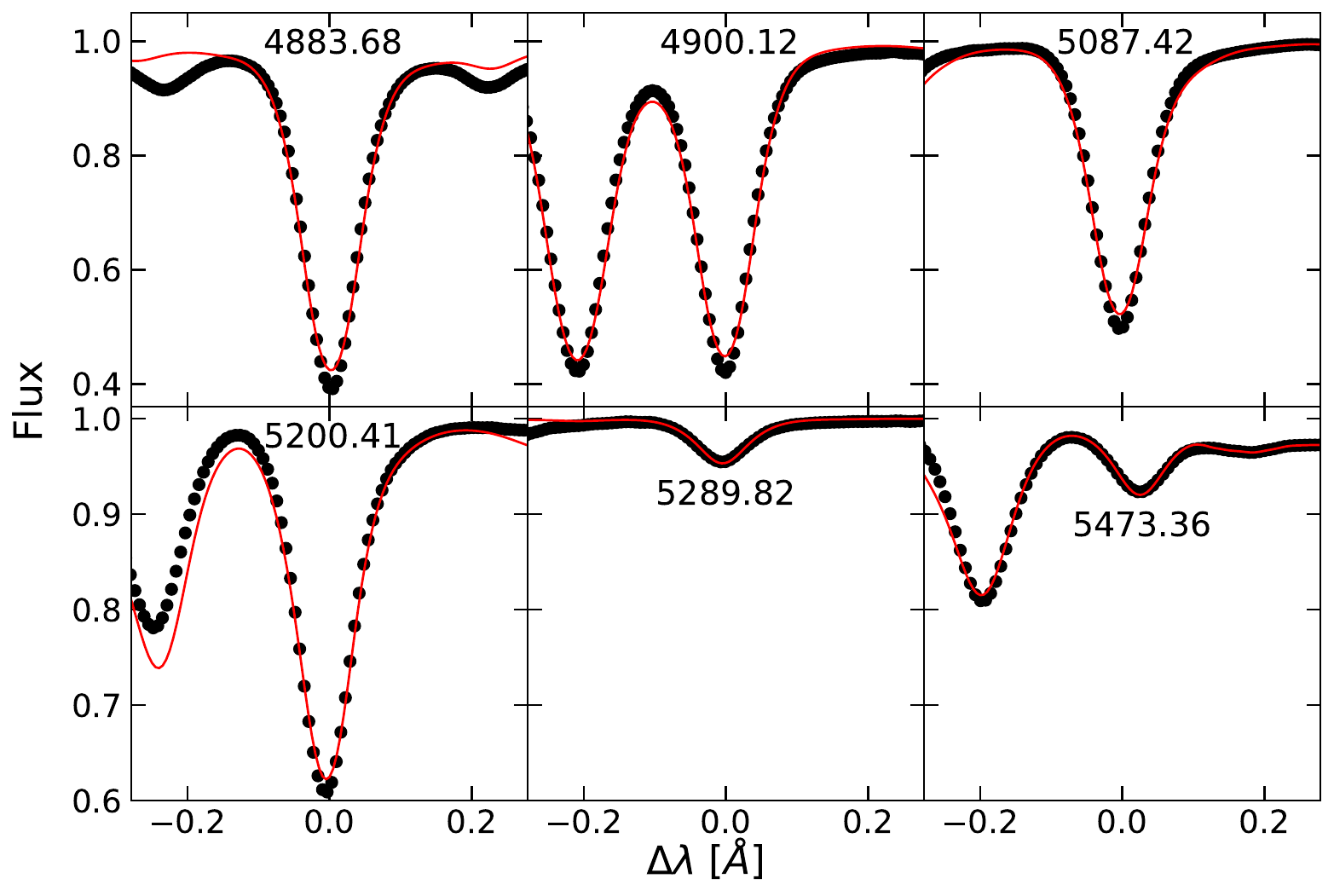}
\caption{Fitted solar spectra for Y II lines with $\vmac = 3$ km/s and $\vmic = 1$ km/s. First line is the diagnostic line 4883.68. The last 5 lines use $\log gf$ values from \citet{Grevesse2015} for comparison.}
\label{fig:solar_y2_grev_loggf}
\end{figure}

\citet{Grevesse2015} carried out a comprehensive study of the abundances of neutron-capture elements in the solar photosphere, including Y in LTE. They used nine Y II lines (see Table \ref{tab:nlte}). For most of these, $\log gf$ from \citet{Hannaford1982} was used. However, lines 4900, 5473 and 5728 were updated based on the lifetimes from \citet{Wannstrom1988}, and \citet{Biemont2011}. By carefully inspecting the line profiles, we found that four of those lines (4125, 4398, 5119, 5728) were very weak or were severely affected by blends. Therefore we limited the abundance analysis to the 5 Y II features (Table \ref{tab:nlte}).

In Fig. \ref{fig:solar_y2_grev_loggf} we show the observed solar Kitt Peak National Observatory (KPNO) FTS atlas \citep{Kurucz1984} in comparison with the best-fit NLTE Y II models. Here we assumed a constant macro-turbulence $\vmac = 3$ km/s and a micro-turbulence $\vmic = 1$ km/s from \citep{Bergemann2012a}. We also include the diagnostic Y II line at $4883.68$ \AA, as the feature is the primary diagnostic in the analysis of solar analogues. Based on these 5 Y II lines, we obtain the average best-fit solar Y abundance of A(Y)$_\textrm{LTE}$ $ = 2.08 \pm 0.04$ dex and A(Y)$_\textrm{NLTE}$ $ = 2.12 \pm 0.04$ dex. The values are significantly different, if we use the $\log gf$ values adopted by \citet{Grevesse2015}, which are mainly based on older transition probabilities from \citet{Hannaford1982} although re-normalised to other sources of lifetimes for some of the lines. In this case, we obtain A(Y)$_\textrm{LTE}$ $ = 2.19 \pm 0.05$ dex and A(Y)$_\textrm{NLTE}$ $ = 2.22 \pm 0.05$ dex, which are also in agreement with the MARCS-based value A(Y)$_\textrm{LTE}$ $ = 2.14 \pm 0.06$ dex by \citet{Grevesse2015}. 

{Currently, there seems to be no strong reason in favour of older $gf$-values, which were adopted by \citet{Grevesse2015}, see also a comprehensive discussion of the improvements in \citealt{Biemont2011}). Hence it is likely that the fact that our photospheric abundance of Y based on \citet{Biemont2011} atomic data is slightly lower compared to the meteoritic value, $2.17 \pm 0.04$ dex \citep{Lodders2009}, is due to the lack of 3D effects in our spectrum synthesis calculations.} Indeed, if we include the 3D abundance corrections obtained by comparing the 3D LTE with 1D LTE estimates from \citet{Grevesse2015}, which amount to $+0.07$ dex (in the sense that 3D abundance are higher compared to 1D), then we would obtain A(Y)$_\textrm{LTE}$ $ = 2.15 \pm 0.08$ dex and A(Y)$_\textrm{NLTE}$ $ = 2.19 \pm 0.08$ dex, in agreement with meteoritic values, however, with a somewhat larger scatter compared to our 1D estimates. The {much increased error may} imply that combining the solutions obtained 1D LTE, 1D NLTE, and 3D LTE is not a reliable approach, and full 3D NLTE calculations would be necessary to further constrain the solar Y abundance, which we defer to a subsequent work in the series.

\begin{figure}
\includegraphics[width=1.00\columnwidth]{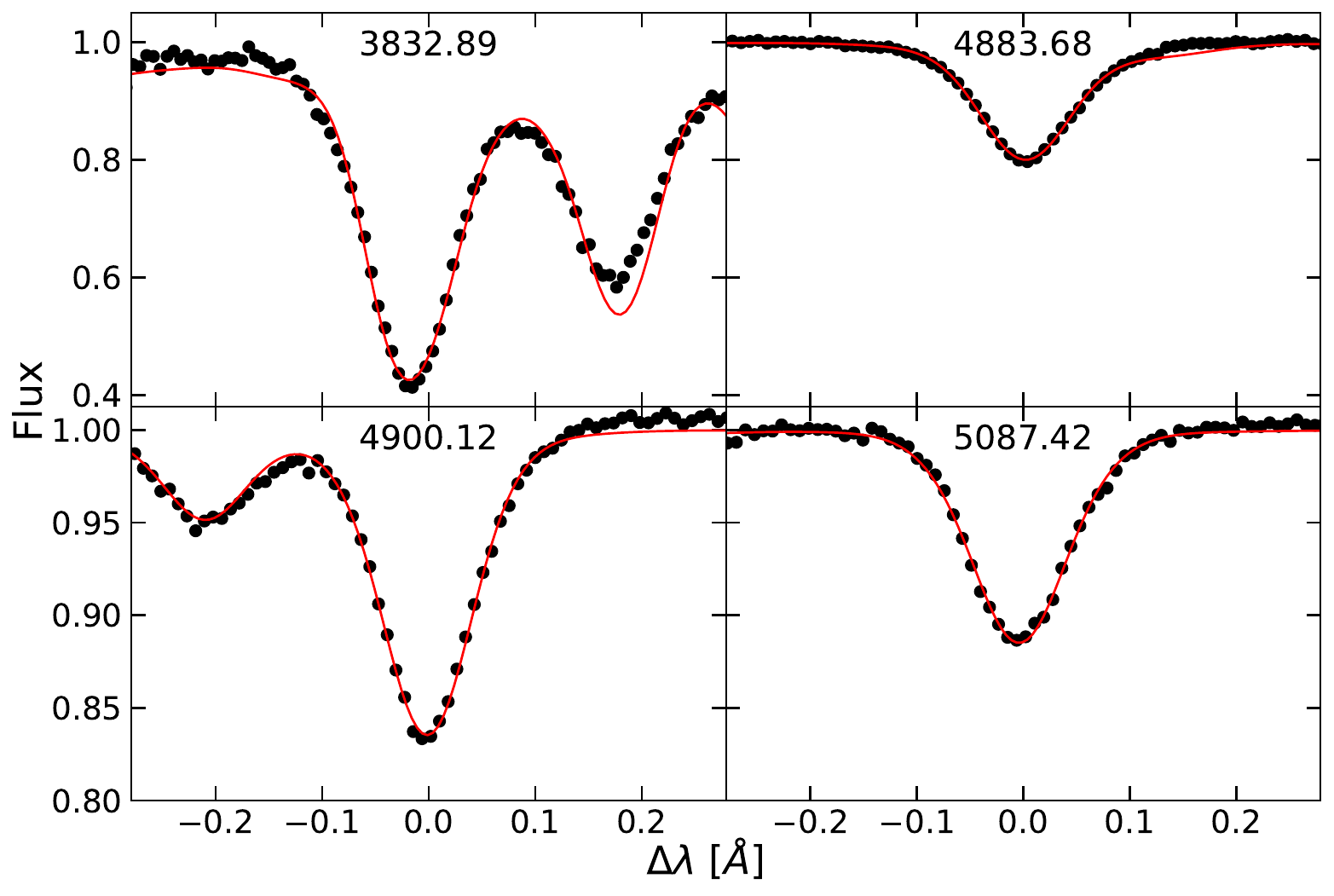}
\caption{HD 122563 fits with one low EP and three medium EP lines using the observed ESPRESSO (R = 190000) spectra.}
\label{fig:hd_fits}
\end{figure}
\begin{figure}

\includegraphics[width=1\columnwidth]{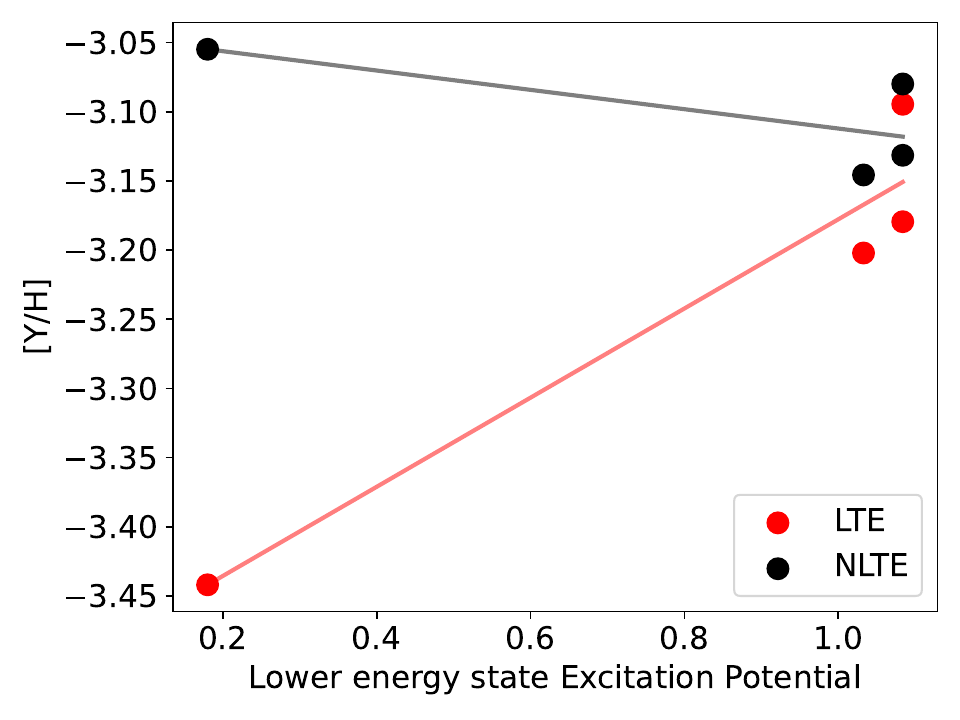}
\includegraphics[width=1\columnwidth]{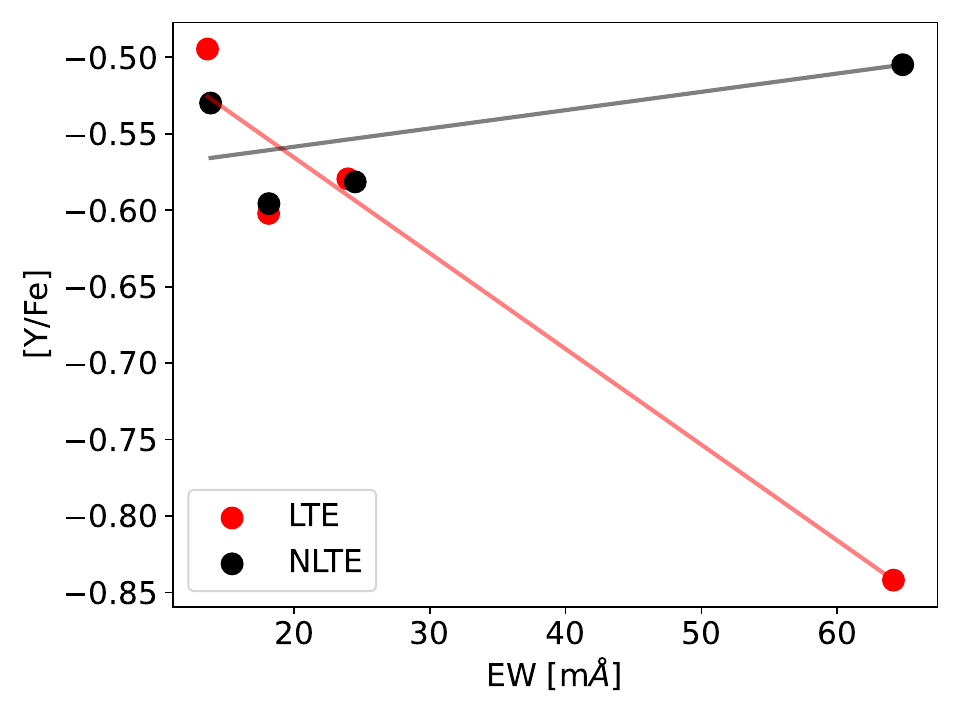}

\caption{Y abundances for HD 122563 ([Y/H] on the left and [Y/Fe] on the right) plotted as a function of lower energy state EP or EW. The red and black points represent respectively fitted LTE and NLTE abundances. Best-fit lines going through the low-EP line 3832.89 are also plotted in relevant colours.}
 \label{fig:yfe_yh_elow}
\end{figure}
\subsection{HD 122563}

An interesting test of the NLTE model atom is to perform the analysis of the excitation-ionisation balance of an element. This test is particularly relevant in the regime, where the NLTE corrections show a different sign and/or amplitude \citep[e.g.][]{Bergemann2012c}. However, no suitable Y I lines are available spectra of FGK type stars, also very few observed spectra have the quality needed to reliably isolate Y II lines of different multiplets.

To test the excitation balance of Y, we have opted for the analysis of the very high-resolution (R = $190\,000$) and SNR ($660.8$) ESPRESSO spectrum \footnote{ESPRESSO data taken within the ESO programme id:0103.D-0118 (PI Adibekyan, pub. date 2021-04-29)} of the well-studied metal-poor red-giant HD 122563. The wavelength coverage extends to 3800 \AA~that allows probing the low- and medium-excitation potential Y II lines. Based on \citet{Bergemann2012a}, \citet{Afsar2016}, \citet{Creevey2019}, and \citet{Sneden2021}, we chose the following stellar parameters for the fit $\teff = 4600$K, $\log g = 1.39$ dex, [Fe/H]$_\textrm{LTE} = -2.60$ dex and [Fe/H]$_\textrm{NLTE} = -2.55$ dex. Here the surface gravity is taken from the asteroseismic analysis by \citet{Creevey2019}. The $\teff$ is in excellent agreement with the interferometric value \citep{Karovicova2020}, $\teff = 4635 \pm 34$ K. {We adopt $\vmic$ of $2.2$ km/s for the analysis based on \citet{Afsar2016,Sneden2021}.}

The best-fit models are compared with ESPRESSO observations in Fig. \ref{fig:hd_fits}. Out of all low-EP lines in Table \ref{tab:nlte}, only the feature at 3832 \AA~was of a sufficient quality to extract a reliable abundance measurement. Of the medium-EP lines, the three features at 4883 \AA, 4900 \AA, and 5087 \AA~are suitable. Owing to the exquisite data quality, the internal error of the fit is very small. Clearly the models yield a good agreement with the data, although we emphasise that the quality of the fit is equally good in LTE and in NLTE.

In Fig. \ref{fig:yfe_yh_elow} ({top} panel), we show the excitation balance of Y II, with the abundance of Y derived from each the four diagnostic Y II features a function of the excitation potential of the lower energy level of each transition. The [Y/H] ratio is useful because it accounts for the differences in the adopted [Fe/H], that is LTE or NLTE. {The possibly most interesting feature of this figure is that NLTE significantly improves the excitation balance of Y II. Specifically, in LTE, we find a large difference between the medium-EP and the low-EP lines, with the latter yielding a $0.28$ dex lower Y abundance and thus A(Y)$_\textrm{LTE}$ $ = -1.02 \pm 0.13$ dex. In NLTE, remarkably, the difference reduces to $0.06$ dex and the low-EP line now yields a slightly higher abundance compared to the medium-EP features. Thus  we obtain a nearly perfect excitation balance in NLTE, with the average abundance of A(Y)$_\textrm{NLTE}$ $ = -0.89 \pm 0.04$ dex, or [Y/Fe] $= -0.55 \pm 0.04$ dex.} For comparison, literature values for HD 122563 include \citet{Aoki2005b}, who obtained A(Y)$_\textrm{LTE}$ $ = -0.84 \pm 0.24$ dex, and \citet{Honda2006} with A(Y)$_\textrm{LTE}$ $ = -0.93 \pm 0.09$ dex. These measurements are in full agreement with our LTE value.

Finally, in Fig. \ref{fig:yfe_yh_elow} ({bottom} panel) we also illustrate the results in the plane of line EW - [Y/Fe], where the latter is the parameters of a primary interest in GCE studies \citep{Travaglio2004, Prantzos2018, Kobayashi2020}. Whereas the figure is conceptually identical to the other panel, it is useful, because it highlights the bias incurred by LTE assumption in the GCE modelling of [Y/Fe]. Also in this figure, for consistency, we use LTE value of [Fe/H] for LTE A(Y) and NLTE value of metallicity for the NLTE estimate of A(Y). Most relevant, in this respect, is the vastly under-estimated [Y/Fe] abundance ratio obtained in LTE from the near-UV low-EP line. {The LTE [Y/Fe] value obtained from this feature is $-0.84$ dex, which is $0.28$ dex lower compared to the LTE value obtained from the optical Y II lines. As we show the stellar parameters of HD 122563 are very well-constrained, with $\teff$ known to $35$ K \citep{Karovicova2020} and $\log g$ to $0.01$ dex \citep{Creevey2019}}. This implies that Y abundance studies that rely on medium-resolution and/or low-SNR spectra, which do not allow for a reliable diagnostics of weak spectral features, may contain a strong systematic bias. Fortunately, the medium-EP optical Y II lines are not affected by this problem, so that rather reliable results can also be obtained from LTE measurements. However, this requires spectra with a very high SNR and high resolving power.

In another example, \citet{Sneden2003} investigated a metal-poor red giant star CS 22892-052  ($\teff = 4700$ K, $\log g$ = 1.0, and [Fe/H] $=-3.10$ dex), using 12 low-EP and 3 medium-EP lines of Y II. They reported an [Y/Fe] ratio lower by $0.15$ to $0.30$ dex compared to that of Sr and Zr, and noting that Y was deviating significantly from the scaled solar system abundance distribution for neutron-capture species from \citet{Burris2000}. From the analysis of our data presented in Fig. \ref{fig:nltecor}, we find the average NLTE correction for the stellar parameters and line selection from \citet{Sneden2003} would amount to $\sim +0.25$ dex, thereby bringing the abundance of Y into agreement with the other first peak s-process elements. Hence, NLTE effects in Y could potentially be relevant within the context of the problem of a "significant deviation from solar neutron-process curve for some lighter neutron-capture elements", as described by these authors.

In summary, we show that NLTE resolves the problem of excitation imbalance  between different Y II lines in the spectrum of a metal-poor red giant HD 122563. Whereas medium excitation potential Y II lines, specifically the optical features at 4883, 4900, and 5087 \AA, can be reliably modelled in LTE, we find that the low excitation-potential lines, such as the near-UV Y II line at 3832 \AA~\citep[e.g.][]{Sneden2003, Lai2008, Cohen2013, Tanriverdi2016}, require NLTE modelling for an unbiased estimate of Y abundance in the spectra of metal-poor stars.
\subsection{Solar analogues and the [Y/Mg] chemical clock}

\begin{figure}
    \centering
    \includegraphics[width=0.48\textwidth]{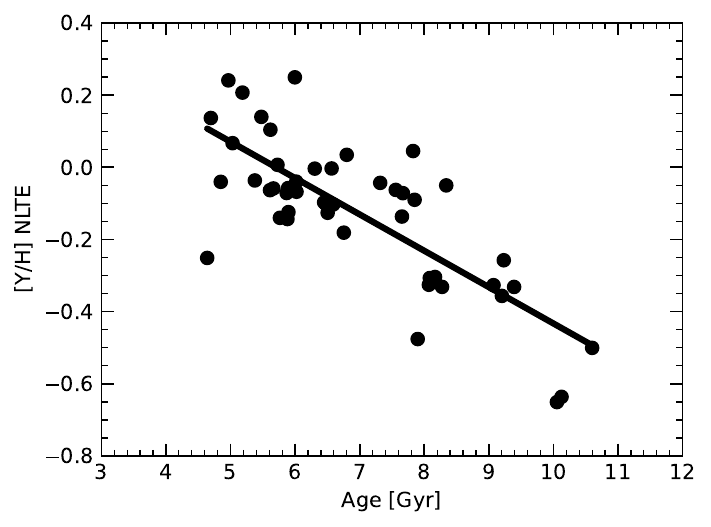}
    \caption{[Y/H] NLTE as a function of age for the sample of stars.}
    \label{fig:yh_age_nlte}
\end{figure}

Finally we explore the consequence of NLTE effects in the Y lines on the use of this element as a Galactic chemical clock. Fig. \ref{fig:yh_age_nlte} shows our NLTE [Y/H] ratios as a function of stellar ages for our sample of solar analogues. To remind, these stars are selected strictly on $\teff$ and $\logg$ to be similar to the Sun, but the [Fe/H] values are unconstrained. The correlation is strong, suggesting that the Y abundance increases with decreasing stellar age. The solution of the best-fit line is given by [Y/H] $ = 0.574 - 0.101 (\pm 0.012) \times$ age [Gyr] with r $= -0.77$, where r is Pearson product-moment correlation coefficient.

\begin{figure}
    \centering
    \includegraphics[width=0.48\textwidth]{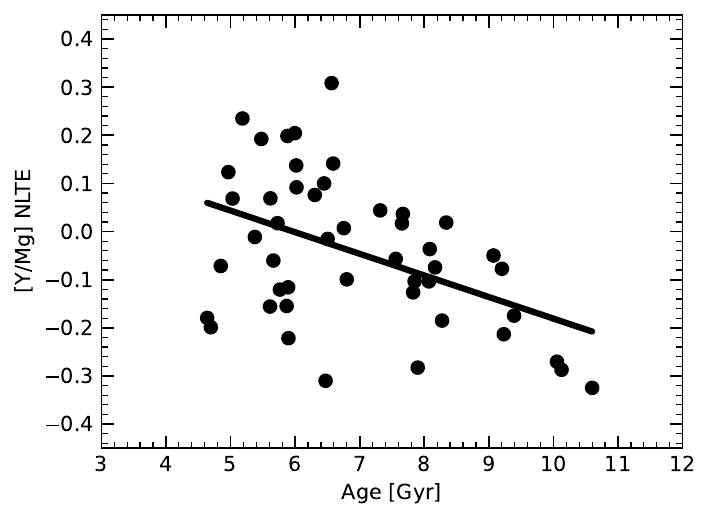}
    \caption{[Y/Mg] NLTE for the sample of stars as a function of age. Best fit least-squared linear fit is also plotted with the equation [Y/Mg] $ = 0.268 - 0.045 (\pm 0.013) \times$ age [Gyr] and r $ = -0.46$.}
    \label{fig:ymg_age_nlte}
\end{figure}

Now we can also investigate how the [Y/Mg] ratio - the 'chemical clock' - behaves as a function of age. For LTE values of [Y/Mg], the correlation with stellar age is weak with r $= -0.48$ and the resulting least-square fit yields [Y/Mg] $ = 0.320 - 0.051 (\pm 0.014) \times$ age [Gyr]. The slope is noticeably higher compared to other studies \citep[such as slope of $- 0.0404 \pm 0.0019$ dex Gyr$^{-1}$ in][]{Nissen2015}. The weak correlation is not surprising, given that our stellar sample probes a substantial spatial volume and their chemical elemental enrichment would also vary. Fig. \ref{fig:ymg_age_nlte} shows the NLTE [Y/Mg] ratios as a function of age, with Mg determined self-consistently in NLTE. {While the core objective of this study is not to directly investigate the Y abundance as a 'chemical clock', but to assess the potential impact of NLTE on this quantity, it is still important to comment on the increased scatter associated with the observed data. As outlined in Sect. \ref{sec:abund_determination}, systematic uncertainties of abundances derived from stellar parameters are larger for Mg than for Y, possibly leading to the increased scatter observed in the [Y/Mg]-age plane. Nevertheless, the combined [Y/Mg] systematic uncertainty is around $0.13$ dex, which is not large enough to explain the [Y/Mg] scatter of approximately $0.20$ to $0.30$ dex. Considering the age uncertainties of 30\%, it is unclear if the total scatter originates from any physical factors or from combined uncertainties. However, as our focus is primarily on the NLTE effect on this relation, we concentrate our analysis on this specific question, and encourage more detailed studies of the [Y/Mg]-age correlation using ages obtained using other complementary quantities, such as asteroseismic measurements.}

The combined NLTE corrections for [Y/Mg] do not exceed $0.08$ dex for all stars in the sample. Thus the equation has a similar although a bit smaller slope with equation [Y/Mg] $ = 0.268 - 0.045 (\pm 0.013) \times$ age [Gyr] and a similar r $ = -0.46$. This would imply that NLTE effects do not significantly influence the [Y/Mg] ratio for stars with $\teff$ and $\logg$  similar to the Sun. Therefore, the applicability of this ratio as a 'chemical clock' for Sun-like stars would not be substantially biased by neglecting NLTE effects.

\section{Conclusions}\label{sec:conclusion}

In this study, we perform a detailed analysis of NLTE effects in the diagnostic lines of Y II that are typically used in the chemical abundance calculations for late-type stars. We develop a new NLTE model atom of Y and explore the sensitivity of the resulting NLTE effects to the still poorly-known collisional data. 

We find that NLTE effects in the Y II lines depend on stellar parameters and on the atomic properties of lines, however, the NLTE corrections are robust with respect to assumptions on H and e$-$ collisions. For the solar photosphere, we obtain the NLTE Y abundance of A(Y)$_{\rm NLTE} = 2.12 \pm 0.04$ dex, which is $0.04$ dex higher compared to the LTE value. The absolute value is slightly lower compared to the abundance of Y in C I chondrites, however, this difference could be either caused by the quality of transition probabilities and/or by 3D effects. Applying the fiducial 3D corrections from \citet{Grevesse2015} to our NLTE value, we obtain a solar photospheric Y abundance A(Y)$_{\rm NLTE} = 2.19 \pm 0.08$ dex consistent with meteorites, however, with a double line-to-line scatter, which may indicate the need for full 3D NLTE calculations to put the solar Y abundance on a firmer ground.

For unevolved - main-sequence and subgiant - stars with metallicities in the range $-4 \lesssim$ [Fe/H] $\lesssim +0.5$, the NLTE corrections are within $+0.2$ dex for all diagnostic Y II lines and typically, for the parameter space of the Galactic disc stars they do not exceed $\pm 0.1$ dex. Especially, medium excitation potential lines of multiplets 21 and 23 are almost unaffected by NLTE. Since these lines form the main Y diagnostics in optical high-resolution spectra of FGK-type stars \citep[e.g.][]{Bensby2014,Nissen2015,TucciMaia2016}, we conclude that no significant biases are associated with LTE-based abundance derived from these features, as long as one does not aim at Y abundances precise to better than $\approx 0.1$ dex. 

For red giants, the NLTE effects are more severe and may exceed $0.5$ dex for the low-excitation potential Y II lines at low metallicities, [Fe/H] $\lesssim -3$ dex. Such spectral lines are typically used in the abundance studies of very- and ultra-metal-poor stars in the Galactic halo \citep[e.g.][]{Johnson2002,Sneden2003,Roederer2014}, indicating that the LTE approximation is not suitable in this regime and may lead to substantial biases in [Y/Fe] abundance ratios. For the Gaia/4MOST benchmark star, HD 122563, we find a significant excitation imbalance in LTE,  with [Y/Fe]$_{\rm LTE}$ $= -0.60~\pm~0.11$ dex, and the lower-excitation potential line yielding a much lower Y abundance compared to medium-excitation potential features. The problem is resolved in NLTE, with [Y/Fe]$_{\rm NLTE}$ $= -0.55~\pm~0.04$ dex, and a very good agreement between all diagnostic Y II lines.

Finally, we show, through a detailed LTE and NLTE abundance analysis of 48 solar analogues with high-quality Gaia-ESO spectra, that the [Y/Mg] ratio in Sun-like stars is not significantly affected by NLTE, confirming the reliability of [Y/Fe] ratio as a potential Galactic chemical clock, at least from the perspective of line formation of Y and Mg. Similar to previous literature studies \citep[e.g.][]{daSilva2012,Nissen2015,Feltzing2017,Jofre2020}, we also find a significant correlation between [Y/Fe] and stellar ages, however, the correlation of [Y/Mg] with ages is far less pronounced, which could possibly be due to less strict limits on our definition of solar analogues, such as the lack of [Fe/H] as a selection criterion \citep[see also][]{Feltzing2017}, or a larger Galactic  volume probed by our stellar sample.

\section*{Acknowledgements}

Based on observations made with ESO Telescopes at the La Silla or Paranal Observatories under programme ID(s) 072.D-0019(B), 072.D-0309(A), 072.D-0337(A), 072.D-0406(A), 072.D-0507(A), 072.D-0742(A), 072.D-0777(A), 073.C-0251(B), 073.C-0251(C), 073.C-0251(D), 073.C-0251(E), 073.C-0251(F), 073.D-0100(A), 073.D-0211(A), 073.D-0550(A), 073.D-0695(A), 073.D-0760(A), 074.D-0571(A), 075.C-0245(A), 075.C-0245(C), 075.C-0245(D), 075.C-0245(E), 075.C-0245(F), 075.C-0256(A), 075.D-0492(A), 076.B-0263(A), 076.D-0220(A), 077.C-0655(A), 077.D-0246(A), 077.D-0484(A), 078.D-0825(A), 078.D-0825(B), 078.D-0825(C), 079.B-0721(A), 079.D-0178(A), 079.D-0645(A), 079.D-0674(A), 079.D-0674(B), 079.D-0674(C), 079.D-0825(B), 079.D-0825(C), 079.D-0825(D), 080.B-0489(A), 080.B-0784(A), 080.C-0718(A), 081.D-0253(A), 081.D-0287(A), 082.D-0726(A), 083.B-0083(A), 083.D-0208(A), 083.D-0671(A), 083.D-0682(A), 083.D-0798(B), 084.D-0470(A), 084.D-0693(A), 084.D-0933(A), 085.D-0205(A), 086.D-0141(A), 087.D-0203(B), 087.D-0230(A), 087.D-0276(A), 088.B-0403(A), 088.B-0492(A), 088.C-0239(A), 088.D-0026(A), 088.D-0026(B), 088.D-0026(C), 088.D-0026(D), 088.D-0045(A), 089.D-0038(A), 089.D-0298(A), 089.D-0579(A), 090.D-0487(A), 091.D-0427(A), 092.D-0171(C), 092.D-0477(A), 093.D-0286(A), 093.D-0818(A), 094.D-0363(A), 094.D-0455(A), 171.D-0237(A), 187.B-0909(A), 188.B-3002(A), 188.B-3002(B), 188.B-3002(C), 188.B-3

Based on observations collected at the European Southern Observatory under ESO programme 0103.D-0118 and data obtained from the ESO Science Archive Facility with DOI(s) under https://doi.org/10.18727/archive/21. 

This work has made use of the VALD database, operated at Uppsala University, the Institute of Astronomy RAS in Moscow, and the University of Vienna.

NS and MB acknowledge funding from the European Research Council (ERC) under the European Union’s Horizon 2020 research and innovation programme (Grant agreement No. 949173).

MB is supported through the Lise Meitner grant from the Max Planck Society. We acknowledge support by the Collaborative Research centre SFB 881 (projects A5, A10), Heidelberg University, of the Deutsche Forschungsgemeinschaft (DFG, German Research Foundation). 

{We would like to express our appreciation to the anonymous reviewer for their insightful comments and suggestions, which significantly contributed to improving the quality of this paper.}
\section*{Data availability}

All work presented in this study relies on \href{http://archive.eso.org/scienceportal/home}{publicly available data acquired within the Gaia-ESO survey}. We also make use of the publicly available \href{https://archive.eso.org/dataset/ADP.2021-04-19T12:45:33.539}{ESPRESSO spectrum of HD 122563}.

\bibliographystyle{mnras}
\bibliography{references}

\appendix

\section{Table with Y II lines}

\begin{table}
\begin{minipage}{\linewidth}
\renewcommand{\footnoterule}{} 
\setlength{\tabcolsep}{3pt}
\caption{Y II lines by multiplets. The standard diagnostic solar lines from \citet[][]{Grevesse2015} are marked with an asterisk $^*$ in the wavelength column.}
\label{tab:nlte}     
\begin{center}
\begin{tabular}{l cccccc}
\noalign{\smallskip}\hline\noalign{\smallskip}  
 Multiplet & $\lambda$ & $\Elow$ & $\Eup$ &  $\log gf$  & Ref.\footnote{References for $\log gf$: ~~~
(1) \citet{Hannaford1982}
(2) \citet{Biemont2011}
(3) \citet{Kurucz2011}
} & Ref. ex. studies\footnote{Reference examples of studies ~~~
(4) \citet{Burris2000}
(5) \citet{Fulbright2000}
(6) \citet{Johnson2002}
(7) \citet{Sneden2003}
(8) \citet{Aoki2005a}
(9) \citet{Ivans2006}
(10) \citet{Lai2008}
(11) \citet{Roederer2010}
(12) \citet{Cohen2013}
(13) \citet{Roederer2014a}
(14) \citet{Bensby2014}
(15) \citet{Nissen2015}
(16) \citet{TucciMaia2016}
(17) \citet{Tanriverdi2016}
(18) \citet{Aoki2017}
(19) \citet{Carrillo2022}

} \\
  & [\AA] & [eV] & [eV] &  \\
\noalign{\smallskip}\hline\noalign{\smallskip}

  &  &  &  &  \\
Mult. 6         &  &  &  & \\
a~$^3$D  - z~$^3$P$*$ &  &  &  & \\
        & 4398.008$^{*}$ & 0.13 & 2.95 & -0.90 & 1 & 4-7, 9, 10, 18\\
Mult. 7          &  &  &  & \\
a~$^3$D  - z~$^1$D$*$  &  &  &  & \\
        & 3950.349  &  0.10 & 3.24 & -0.49 & 1 & 6-8, 10, 18\\
Mult. 8         &  &  &  &  \\
a~$^3$D  - z~$^3$F$*$ &  &  &  &  \\
        & 3710.287 &  0.18 & 3.52 & 0.46 & 1 & 6, 9, 10 \\  
        & 3774.330 &  0.13 & 3.41 & 0.21  & 1 & 6-9, 10, 13\\  
        & 3788.693 &  0.10 & 3.76 & -0.07 & 1 & 6-8, 10\\
        & 3818.341 &  0.13 & 3.76 & -0.98 & 1 & 7, 8, 10\\
        & 3832.889 &  0.18 & 3.41 & -0.34  & 1 & 7, 10, 12, 17\\
Mult. 9         &  &  &  &  \\
a~$^3$D  - z~$^1$P$*$ &  &  &  &  \\
        & 3747.551 & 0.10 & 3.41 & -0.91  & 1 & 6, 7, 10\\ 
Mult. 21         &  &  &  & \\
a~$^3$F  - z~$^3$F$*$  &  &  &  & \\
       & 5087.418$^{*}$ & 1.08 & 3.52 & -0.16 & 2 & 5-9, 11, 13-15, 19\\
       & 5200.406$^{*}$ & 0.99 & 3.38 & -0.47 & 2 & 5, 7, 9, 11, 13-16, 19\\
       & 5205.722 & 1.03 & 3.41 & -0.28 & 2 & 5-7, 9, 13, 19\\
       & 5289.815$^{*}$ & 1.03 & 3.38 & -1.68 & 3 & 9, 19\\
Mult. 23         &  &  &  & \\
a~$^3$F  - z~$^3$D$*$  &  &  &  & \\
        & 4854.861  & 0.99 & 3.55 & -0.27 & 2 & 6, 9, 10, 14, 16\\
        & 4883.682  & 1.08 & 3.62 & 0.19 & 2 & 5, 6, 8-10, 13-15\\
        & 4900.118$^{*}$  & 1.03 & 3.56 & 0.03 & 2 & 5, 6, 9, 10, 14\\
Mult. 26         &  &  &  & \\
a~$^3$P  - y~$^3$P$*$  &  &  &  & \\
        & 5473.385$^{*}$  & 1.74 & 4.00 & -0.78 & 2 & 9\\
Mult. 33         &  &  &  & \\
b~$^1$D  - z~$^1$F$*$  &  &  &  & \\
        & 5402.773 &  1.84 & 4.13 & -0.31 & 2 & 14, 16\\
Mult. 35         &  &  &  & \\
a~$^1$G  - z~$^1$F$*$  &  &  &  & \\
        & 5662.922 &  1.94 & 4.13 & 0.34 & 2 & 14\\

\noalign{\smallskip}\hline\noalign{\smallskip}
\end{tabular}
\end{center}
\end{minipage}
\end{table}

\bsp
\label{lastpage}
\end{document}